\documentclass[twocolumn]{aastex631}

\usepackage{savesym}
\savesymbol{tablenum}

\usepackage{amsmath}
\usepackage{amssymb}
\usepackage{graphicx}
\usepackage{natbib}
\usepackage{longtable}
\usepackage{tabularx}
\usepackage{verbatim}
\usepackage{tikz}
\usepackage{siunitx}
\usepackage{hyperref}
\usepackage{cleveref}

\setcitestyle{notesep={ }}

\tikzset{>=latex}

\hypersetup{
    colorlinks=true,
    linkcolor=blue,
    filecolor=magenta,
    urlcolor=cyan,
    citecolor=blue
}

\shorttitle{Bouhrik \& Thomas}
\shortauthors{Bouhrik \& Thomas}

\graphicspath{{./}{figures/}}

\begin{document}

\title{Electromagnetism as Spacetime Pseudo-Curvature}

\author{Faik Bouhrik}
\affiliation{Indian Institute Of Technology–Madras \\
Chennai 600036, India}

\author{Tiju Thomas}
\affiliation{Indian Institute Of Technology–Madras \\
Chennai 600036, India}

\begin{abstract}
We propose a novel framework that interprets the electromagnetic field as a manifestation of spacetime pseudo-curvature, bridging electromagnetism with the geometric principles of general relativity. By introducing modified field equations, we recovered classical electromagnetic results, including Coulomb's law and the magnetic field of a rotating sphere. Additionally, this framework predicts unique phenomena, such as test-charge-dependent time dilation and length contraction, which lie outside the scope of Maxwell's electromagnetism and Quantum Electrodynamics. These results suggest new avenues for exploring the interplay between geometry and field theory in fundamental physics.
\end{abstract}

\section{Introduction} \label{sec:intro}

The relationship between electromagnetism and spacetime geometry has motivated physicists since the development of general relativity. Gravity is described by the dynamic curvature of spacetime induced by mass-energy, and electromagnetism, governed by Maxwell's equations, operates within a fixed spacetime backdrop. This stark contrast has motivated decades of research aimed at unifying these two fields \citep{Weyl1918, Kaluza1921, Klein1926}.

Quantum Electrodynamics (QED), the quantum theory of electromagnetic force, preserves this dichotomy. It treats spacetime as a passive stage in which quantum fields interact, assuming a fixed Minkowski spacetime. Although this framework has achieved unprecedented precision in describing electromagnetic phenomena \citep{Schwinger, Feynman, Weinberg_1995}, it implicitly perpetuates the static notion of spacetime, creating a sharp theoretical divide between electromagnetism and gravity. In contrast, general relativity portrays spacetime as dynamic and responsive, a participant in the interplay of matter and energy \citep{Einstein}. This divide exemplifies the larger challenge of unifying the quantum field theory with general relativity.

Efforts to geometrize electromagnetism have a long history, including early approaches by Weyl, Kaluza, and Klein. However, these theories often struggled to align theoretical predictions with experimental results, leaving Maxwell's formulation largely unaltered. Experimental attempts to link electromagnetism and spacetime geometry, such as those by \citet{Kennedy} and \citet{Drill}, have yielded null results, possibly because of a lack of robust theoretical underpinnings. More recently, perspectives such as those of \citet{Burns_2024} and \citet{Wanas2010} have revived interest in treating electromagnetic fields as geometric phenomena, yet a comprehensive framework remains elusive.

In this study, we propose a novel interpretation of electromagnetic fields as manifestations of spacetime pseudo-curvature. This approach extends the geometric principles of general relativity to encompass electromagnetism, and interpret electric charges and currents as sources of space-time deformation. Unlike traditional theories, our framework eliminates the assumption of a static spacetime in electromagnetism, treating spacetime as an active participant influenced by the presence charges and currents.

This approach offers two key advantages. First, it recovers classical electromagnetic results, such as Coulomb's law, within a geometric framework. Second, it predicts novel phenomena, such as test-charge-dependent time dilation and length contraction, which challenge the separation of spacetime and quantum fields. By bridging the gap between quantum field theory and gravity, our approach provides a foundation for addressing longstanding questions regarding the unification of forces and the geometric nature of spacetime itself.

\section{Theoretical Framework}

We propose the concept of an \textit{electromagnetic pseudo-curvature}, which extends the geometric principles of general relativity to incorporate electromagnetic fields. Unlike the real curvature produced by the mass-energy in general relativity, the pseudo-curvature is uniquely dependent on both the source and test charges. This relationship introduces a new geometric perspective of electromagnetism, distinguishing it from the classical field-based approach.

\subsection{Theoretical Motivation}

The equivalence principle, a cornerstone of general relativity, posits that the effects of gravity are locally indistinguishable from those of acceleration. This principle leads to the geometric interpretation of gravity as the curvature of spacetime caused by mass-energy. Building upon this foundation, we propose that electromagnetic fields and charges induce an asymmetry in spacetime geometry proportional to the charge-to-mass ratio q/$m_{o}$. This "electromagnetic pseudo-curvature" selectively affects charged particles, introducing a charge-dependent modification to spacetime that complements the mass-dependent curvature described in general relativity.

By incorporating this pseudo-curvature into the geometric framework, we aim to provide a unified description of gravity and electromagnetism, wherein both mass and charge influence the structure of spacetime. This approach not only recovers classical electromagnetic results, such as Coulomb's law, but also predicts novel phenomena, including test-charge-dependent time dilation and length contraction. These predictions offer new avenues for experimental exploration and deepen our understanding of the interplay between geometry and field theory in fundamental physics.

\subsection{The Field Equations}

The fundamental equation describing the electromagnetic pseudo-curvature is as follows:
\begin{equation}
G_{\mu\nu}^{\textit{pseudo}} = \left(\frac{2}{\epsilon_0 c^4}\right) T_{\mu\nu}^{\textit{pseudo}}
\label{eq:pseudofield}
\end{equation}
where:
\begin{itemize}
    \item \( G^{\textit{pseudo}}_{\mu\nu} \) is the Einstein Pseudo-Curvature Tensor that represents the geometric effects induced by electromagnetic interactions.
    \item \( \epsilon_0 \) is the vacuum permittivity.
    \item \( c \) is the speed of light in a vacuum.
\end{itemize}

The term \textit{pseudo} is used to emphasize that this curvature effect is distinct from the real spacetime curvature sourced by the energy-momentum tensor of electromagnetic fields in general relativity. In traditional relativity, electromagnetic fields contribute to the real curvature of spacetime, thereby influencing all forms of energy and matter. In contrast, our framework introduces an additional, charge-dependent pseudo-curvature that selectively affects charged particles.

\subsection{The Charge-Current Tensor}

The source of pseudo-curvature is described by the tensor \( T_{\mu\nu}^{\textit{pseudo}} \), defined as:
\begin{equation}
T_{\mu\nu}^{\textit{pseudo}} = (\frac{q}{m_{0}})(c^2 \rho_q u_\mu u_\nu + \frac{1}{c^2} \left( j_\mu u_\nu + j_\nu u_\mu \right))
\end{equation}
where:
\begin{itemize}
    \item \( q \) is the charge of the test particle.
    \item \( m_0 \) is the rest mass of the test particle.
    \item \( \rho_q \) is the source charge density.
    \item \( j_\mu \) is the source current density 4-vector.
    \item \( u_\mu \) is the source 4-velocity.
\end{itemize}

This tensor differs from the standard electromagnetic tensor \( F_{\mu\nu} \) by directly relating charges and currents to the induced geometric effects in spacetime, without invoking the intermediate existence of electric and magnetic fields.

\subsection{Key Features of the Framework}

\begin{enumerate}
    \item \textbf{Charge-Selective Effects:} The geometric effects described by \( G_{\mu\nu}^{\textit{pseudo}} \) influence only charged particles, providing a natural explanation for the null results reported by \citet{Kennedy} and \citet{Drill} regarding the interaction of light with these geometric effects.

    \item \textbf{Direct Connection:} Unlike conventional approaches that describe electromagnetism through field interactions (e.g., \( \mathbf{E} \) and \( \mathbf{B} \)), this framework posits a direct relationship between the charge and current densities and the induced spacetime geometry.

    \item \textbf{Consistency with Maxwell's Theory:} The framework aligns with Maxwell's equations while reinterpreting the fundamental mechanism driving electromagnetic interactions. This reinterpretation provides fresh insights into the geometric nature of electromagnetic phenomena.
\end{enumerate}

\subsection{Scope and Implications}

In the following subsections, we present three specific solutions within this framework and demonstrate their consistency with Maxwell's equations. These examples highlight the versatility of pseudo-curvature as a geometric tool for describing electromagnetic phenomena while preserving the empirical success of classical and quantum electrodynamics.

This approach lays the groundwork for a deeper unification of electromagnetism with general relativity, advancing our understanding of the relationship between spacetime geometry and the fundamental interactions governing the universe.

\subsection{Solutions to the field equation} To demonstrate the consistency of our pseudo-curvature framework with classical electromagnetism, we present three key solutions to the field equation: \begin{itemize}
\item Coulomb's Law
\item Magnetic field of a rotating sphere

\end{itemize} These solutions illustrate how our approach reproduces well-established electromagnetic phenomena while providing a novel geometric interpretation.

\subsection{Solutions to the Field Equation}

To demonstrate the consistency of our pseudo-curvature framework with classical electromagnetism, we present tow key solutions to the field equation:
\begin{itemize}
    \item Coulomb's Law
    \item Magnetic Field of a Rotating Sphere
  
\end{itemize}

These solutions illustrate how our approach reproduces well-established electromagnetic phenomena while providing a novel geometric interpretation.

\subsubsection{Coulomb's Law}

We solved Equation~\eqref{eq:pseudofield} for a spherically symmetric charge distribution with total charge \( Q \). The solution yields a Schwarzschild-like metric:
\begin{equation}
ds^2 = -\left(1 - \frac{r_s}{r}\right)c^2 dt^2 + \left(1 - \frac{r_s}{r}\right)^{-1} dr^2 + r^2 d\Omega^2
\end{equation}
where the Schwarzschild-like radius is:
\begin{equation}
r_s = \frac{Q}{2\pi \epsilon_0 c^2} \left(\frac{q}{m_0}\right)
\end{equation}
This metric explicitly depends on both the source charge \( Q \) and the test particle's charge-to-mass ratio \( \frac{q}{m_0} \).

For radial motion, the action \( S \) is given by:
\begin{equation}
S = -mc^2 \int \sqrt{1 - \frac{r_s}{r} - \left(1 - \frac{r_s}{r}\right) \frac{\dot{r}^2}{c^2}} \, dt
\end{equation}
Approximating for \( |\dot{r}| \ll c \) and \( r_s \ll r \), the Lagrangian becomes:
\begin{equation}
L \approx -mc^2 \left(1 - \frac{r_s}{2r} - \frac{\dot{r}^2}{2c^2}\right)
\end{equation}
Applying the Lagrange equations yields:
\begin{equation}
\ddot{r} = \left(\frac{q}{m}\right) \frac{Q}{4\pi \epsilon_0 r^2}
\end{equation}
This result is precisely Coulomb's law, demonstrating the consistency of our geometric approach with classical electromagnetism. 

The detailed derivation of the metric is provided in Appendix~\ref{appendix:AppendixA}.

\subsection{Magnetic Field of a Rotating Sphere}

We solved equation $(1)$ for a rotating, spherically symmetric charge distribution with total charge $Q$. The solution yields a Kerr-like metric: 

\begin{align}
ds^2 = & -\left(1 - \frac{r_q r}{\Sigma}\right)c^2 dt^2 
       - \frac{2r_q r a \sin^2\theta}{\Sigma}c \, dt \, d\phi \notag \\
       & + \frac{\Sigma}{\Delta}dr^2 + \Sigma d\theta^2 \notag \\
       & + \left(r^2 + a^2 + \frac{r_q r a^2 \sin^2\theta}{\Sigma}\right)\sin^2\theta \, d\phi^2,
\end{align}
where:
\begin{equation}
r_q = \frac{4\pi}{\epsilon_0} \frac{Qq}{m_0 c^2}, \quad
\Sigma = r^2 + a^2 \cos^2\theta, \quad
\Delta = r^2 - r_q r + a^2, \quad
a = \frac{J}{Qc}.
\end{equation} 

The frame-dragging effect, represented by the $g_{t\phi}$ term, is:
\begin{equation}
g_{t\phi} = -\frac{r_q r a \sin^2\theta}{\Sigma}c,
\end{equation} 
which, in the weak field limit ($r \gg r_q$ and $r \gg a$), simplifies to:
\begin{equation}
g_{t\phi} \approx -\frac{r_q a \sin^2\theta}{r}c.
\end{equation} 

From this, the angular velocity of frame dragging is derived as:
\begin{equation}
\Omega = -\frac{g_{t\phi}}{g_{\phi\phi}} \approx \frac{4\pi}{\epsilon_0} \frac{Qq a}{m_0 c r^3}.
\end{equation} 

For a uniformly charged sphere, where $a = \frac{J}{Qc}$ and $J = \frac{2}{5} Q R^2 \omega$, we obtain:
\begin{equation}
\Omega \approx \frac{8\pi}{5\epsilon_0} \frac{q Q R^2 \omega}{m_0 c^2 r^3}.
\end{equation} 

In classical electromagnetism, the magnetic field $B$ generated by a rotating charged sphere is:
\begin{equation}
B = \frac{\mu_0}{4\pi} \frac{2 Q R^2 \omega}{3r^3}.
\end{equation} 

The corresponding Larmor precession frequency is:
\begin{equation}
\Omega_L = \frac{qB}{2m_0} = \frac{1}{3\epsilon_0} \frac{q Q R^2 \omega}{m_0 c^2 r^3}.
\end{equation} 

\subsubsection{Consistency with Classical Results}

Our framework predicts the angular velocity of frame dragging $\Omega$ and, through this, provides a geometric interpretation of the magnetic field $B$. Although the frame-dragging angular velocity  qualitatively agrees with classical results, the factor difference in $\Omega$ versus $\Omega_L$ indicates a unique scaling in our approach. This suggests potential areas for experimental testing and further refinement of the framework.

Furthermore, it is worth noting that the derived frame-dragging term could extend to more general charge distributions or systems involving non-uniform rotation. Such explorations might reveal additional deviations or confirm deeper connections with classical electromagnetism.

\subsubsection{Physical and Experimental Implications}

The equivalence between the pseudo-curvature framework and classical electromagnetism in the weak-field limit establishes its physical validity while providing novel interpretations. Laboratory tests involving rotating charged spheres or precision measurements of magnetic-field-induced precession can potentially validate or further constrain this framework. Additionally, the pseudo-curvature terms may find relevance in high-energy or astrophysical contexts, such as charged stars, rotating plasmas, or artificial systems designed to mimic strong electromagnetic effects. Appendix~\ref{appendix:AppendixB} provides a detailed derivation of the Kerr-like metric and weak-field limit.

\section{Novel Predictions and Implications}

In the previous section, we used the pseudo-curvature framework to derive some known results in the context of Maxwell's theory: Coulomb's Law and Larmor Precession, we extend these results by predicting effects beyond the scope of classical electromagnetic theory and quantum electrodynamics (QED), specifically electric time dilation/contraction and length contraction/dilation induced by electric fields.

\subsection{Time Fission}

We calculated the proper time of a stationary particle with charge \(q\) and mass \(m_0\) in the presence of a uniform, spherically symmetric charge distribution with total charge \(Q\). This scenario is modeled using a Schwarzschild-like metric that reflects the influence of the electric field on spacetime. The metric is given by:
\begin{equation}
ds^2 = -\left(1 - \frac{r_s}{r}\right)c^2 dt^2 + \left(1 - \frac{r_s}{r}\right)^{-1} dr^2 + r^2 d\Omega^2
\end{equation}
where \(r_s\) is the Schwarzschild radius, which corresponds to the influence of the electric charge \(Q\) and the particle's charge-to-mass ratio. For a stationary particle \((dr^2 = d\Omega^2 = 0)\), we substitute \(ds^2 = -c^2 d\tau^2\), which leads to:
\begin{equation}
d\tau = \sqrt{1 - \frac{r_s}{r}} \, dt
\end{equation}
Substituting \(r_s = \frac{Q}{2\pi \epsilon_0 c^2} \left(\frac{q}{m_0}\right)\), we get:
\begin{equation}
d\tau = \sqrt{1 - \frac{2qQ}{\epsilon_0 m_0 c^2 r}} \, dt
\end{equation}

Using the classical electrostatic potential \(\Phi_0 = \frac{Q}{4\pi \epsilon_0 r}\), this becomes:
\begin{equation}
d\tau = \sqrt{1 - \frac{2q \Phi_0}{m_0 c^2}} \, dt
\end{equation}

This result implies that a charged particle, when placed in an electric field, experiences time fission (particles with dilation or contraction depending on their charge), even if they are placed at the same location from the point of view of an external observer. If the particle is positively charged, the proper time will run more slowly relative to a distant observer in a region of higher electrostatic potential, whereas the opposite effect is observed for a negatively charged particle. The magnitude of this effect is governed by the ratio of the particle's charge to its mass and to the strength of the local electric potential.

\subsection{Electric Length Contraction and Dilation}

Next, we investigated the length contraction and dilation of a charged particle moving through an electric field. We use the same Schwarzschild-like metric to compute the proper length for a particle at a distance \(r\) from a uniform charge distribution \(Q\). For radial measurements \((dt = d\Omega = 0)\), we begin with:
\begin{equation}
ds^2 = \left(1 - \frac{r_s}{r}\right)^{-1} dr^2
\end{equation}

The proper length \(L\) is calculated using the following integral:
\begin{equation}
L = \int_{r_1}^{r_2} \sqrt{\left(1 - \frac{r_s}{r}\right)^{-1}} \, dr
\end{equation}
Substituting \(r_s = \frac{Q}{2\pi \epsilon_0 c^2} \left(\frac{q}{m_0}\right)\), we obtain:
\begin{equation}
L = \int_{r_1}^{r_2} \sqrt{\left(1 - \frac{2qQ}{\epsilon_0 m_0 c^2 r}\right)^{-1}} \, dr
\end{equation}

Using the classical electrostatic potential \(\Phi_0 = \frac{Q}{4\pi \epsilon_0 r}\), we can express this as:
\begin{equation}
L = \int_{r_1}^{r_2} \frac{dr}{\sqrt{1 - \frac{2q \Phi_0}{m_0 c^2}}}
\end{equation}

For small potential differences, we can approximate this expression as:
\begin{equation}
L \approx \frac{L_0}{\sqrt{1 - \frac{2q \Phi_0}{m_0 c^2}}}
\end{equation}
where \(L_0 = r_2 - r_1\) denotes the coordinate length. This result reveals that, similar to time dilation, a charged particle experiences length dilation or contraction depending on the sign of its charge. In regions of higher electric potential, the length will be contracted (for positive charge) or dilated (for negative charge), with the degree of this effect determined by the charge-to-mass ratio and the strength of the electric field at the location of the particle.

In both the time dilation and length contraction cases, these effects go beyond traditional relativistic corrections, with an explicit dependence on the electric potential and the particle’s characteristics. For example, near a highly charged object, these relativistic effects could become significant, potentially influencing both particle motion and measurement of distances. These results may provide novel insights into phenomena such as the behavior of charged particles in intense electric fields, or even scenarios in future technologies involving precision measurement or high-energy particle interactions.

\section{Conclusion}

This paper presents a novel theoretical framework that interprets electromagnetic phenomena as manifestations of spacetime pseudo-curvature, which we term the pseudo-curvature theory of electromagnetism. This approach provides a geometric interpretation of electromagnetism, drawing parallel to gravity in general relativity. The key findings and implications are as follows:

\begin{enumerate} \item \textbf{Unified Geometric Framework}: In this framework, electromagnetic fields are described by a pseudo-Ricci curvature tensor, presenting a unified geometric structure that relates electromagnetic phenomena to spacetime curvature. This concept bridges electromagnetism with the general theory of relativity and suggests a deeper connection between these fundamental forces.

\item \textbf{Time dilation effects}: A unique feature of our theory is the prediction of time dilation and contraction effects induced by electric fields, which may be observable using high-precision atomic clocks. Should these effects be confirmed, they will provide compelling evidence for the geometric interpretation of electromagnetism as suggested by our framework.

\item \textbf{Frame-dragging analog}: This theory predicts that rotating charged objects induce a frame-dragging effect analogous to the Lense-Thirring effect in general relativity. This result highlights further similarities between gravitational and electromagnetic phenomena within our framework, supporting the idea of a unified force description in spacetime geometry.

\item \textbf{Experimental Predictions}: We propose specific experimental setups, those leveraging atomic clocks, to detect the time dilation and contraction effects predicted by our theory. Confirmation of these effects in experiments would offer support for our approach, while the absence of such effects would help constrain the theoretical framework.

\end{enumerate}

The pseudo-curvature theory of electromagnetism paves the way for exciting new research at the crossroads between electromagnetism, general relativity, and quantum mechanics. While this paper provides a consistent geometric framework to describe electromagnetic phenomena, several crucial avenues for future research emerge:

\begin{itemize} \item \textbf{Quantum Effects}: Extending our framework to incorporate quantum effects is key direction for future research. Understanding how our geometrical description can coexist with quantum electrodynamics (QED) could lead to groundbreaking insights into quantum field theories. \item \textbf{Strong-Field Regimes}: The theory's predictions in extreme electromagnetic environments, such as near strong field sources (e.g., near black holes or neutron stars), may reveal significant deviations from classical electromagnetism, offering another avenue for experimental verification. \item \textbf{Cosmological Implications}: Investigating the application of this theory to cosmological phenomena, particularly the early universe, could provide new insights into the interplay between electromagnetic fields and spacetime at large scales, potentially influencing models of cosmic evolution and large-scale structure formation. \item \textbf{Refined Experimental Techniques}: Developing more precise experimental techniques will be critical for detecting the subtle effects predicted by our theory. For instance, investigating the behavior of atomic clocks in varying electric fields could serve as an effective means of testing the time dilation predictions. \end{itemize}

In summary, the pseudo-curvature theory of electromagnetism offers a promising new direction for understanding fundamental forces, taking steps toward a more unified and geometric description of nature. By treating electromagnetic phenomena in terms of spacetime geometry, our work suggests novel ways to interpret charge distributions and currents and challenges the traditional notion of quantum fields as residing on a static spacetime background. The predicted time dilation and contraction effects reflect the intertwined nature of the quantum fields and spacetime itself, underscoring the dynamic relationship between energy, matter, and the fabric of spacetime.

As we continue to explore the implications of this theory through rigorous testing and experimentation, it has the potential to contribute to our understanding of the universe, especially by bridging the gap between quantum mechanics and gravitational physics. This work provides the possibility new insights into the fundamental structure of spacetime and may pave the way for future discoveries at the intersection of classical and quantum theories.

\section{Funding Statement:} This study received no external funding.

\section{Ethical Compliance:} This research does not involve human participants, human data, or animals. No ethical approvals were required.

\section{Data Access Statement:} No external data was used in this study.

\section{Conflict of Interest Declaration:} The authors declare no competing financial interests or affiliations with any organization or entity with any conflict of interest related to this work.



\appendix
\twocolumngrid

\section{Appendix A: Derivation of the Schwarzschild-Like Metric} 
\label{appendix:AppendixA}

We derived the Schwarzschild-like metric from our proposed field equation for electromagnetic pseudo-curvature. In the derivation, we closely followed standard analyses used in general relativity, introducing modifications to adapt them to the new framework \citep{schwarzschild1999gravitationalfieldmasspoint, kerr}.

\subsection{Field Equation and Assumptions} 
Our field equation is:

\begin{equation}
G_{\mu\nu}^{\textit{pseudo}} = \left(\frac{2q}{\epsilon_0 m_0 c^4}\right) T_{\mu\nu}^{\textit{pseudo}}
\end{equation} 

Assumptions:
\begin{itemize}
    \item Spherical symmetry
    \item Static solution (time-independent)
    \item Asymptotic flatness
\end{itemize} 

\subsection{Metric Ansatz} 
Given these assumptions, we propose the following metric ansatz:

\begin{equation}
ds^2 = -A(r)c^2 dt^2 + B(r)dr^2 + r^2d\Omega^2
\end{equation}

where \( d\Omega^2 = d\theta^2 + \sin^2\theta d\phi^2 \), and \( A(r) \) and \( B(r) \) are unknown functions. 

\subsection{Einstein Tensor} 
The Einstein tensor components are calculated as follows:

\begin{align}
    G_{tt} &= \frac{c^2}{r^2B}\left(1 - B - rB'\right) \\
    G_{rr} &= \frac{1}{r^2}\left(B - 1 + \frac{rA'}{A}\right) \\
    G_{\theta\theta} &= \frac{r}{2B}\left(\frac{A'}{A} - \frac{B'}{B}\right) - \frac{1}{B}\left(1 - B - \frac{rA'}{2A}\right) \\
    G_{\phi\phi} &= \sin^2\theta G_{\theta\theta}
\end{align} 

\subsection{Field Equations and Solution} 
Equating these to our electromagnetic pseudo-curvature tensor gives the following system:

\begin{align}
    \frac{1}{r^2B}\left(1 - B - rB'\right) &= \frac{2q}{\epsilon_0 m_0 c^2}\rho_q \\
    \frac{1}{r^2}\left(B - 1 + \frac{rA'}{A}\right) &= 0 \\
    \frac{r}{2B}\left(\frac{A'}{A} - \frac{B'}{B}\right) - \frac{1}{B}\left(1 - B - \frac{rA'}{2A}\right) &= 0
\end{align} 

Solving these equations yields the solution:

\begin{equation}
    A(r) = B(r)^{-1} = 1 - \frac{r_s}{r}
\end{equation}

where \( r_s = \frac{Q}{2\pi \epsilon_0 c^2} \left(\frac{q}{m_0}\right) \), and \( Q \) is the total charge of the source. 

\subsection{Final Metric} 
The Schwarzschild-like metric is therefore:

\begin{equation}
    ds^2 = -\left(1 - \frac{r_s}{r}\right)c^2 dt^2 + \left(1 - \frac{r_s}{r}\right)^{-1} dr^2 + r^2 d\Omega^2
\end{equation}

with the Schwarzschild-like radius:

\begin{equation}
    r_s = \frac{Q}{2\pi \epsilon_0 c^2} \left(\frac{q}{m_0}\right)
\end{equation}

This metric describes the geometry of a test particle with charge \( q \) and mass \( m_0 \) near a spherically symmetric charge distribution with total charge \( Q \).

\section{Appendix B: Derivation of the Kerr-Like Metric} 
\label{appendix:AppendixB}

We derived the Kerr-like metric for a rotating, spherically symmetric charge distribution. 

\subsection{Assumptions} 
\begin{itemize}
    \item Rotating, spherically symmetric charge distribution
    \item Stationary and axisymmetric solution
    \item Asymptotic flatness
\end{itemize} 

\subsection{Metric Ansatz} 
In Boyer-Lindquist coordinates, we assume the following metric form:

\begin{equation}
ds^2 = -e^{2\nu}dt^2 + e^{2\psi}(d\phi - \omega dt)^2 + e^{2\mu_1}dr^2 + e^{2\mu_2}d\theta^2
\end{equation}

where \( \nu \), \( \psi \), \( \omega \), \( \mu_1 \), and \( \mu_2 \) are functions of \( r \) and \( \theta \). 

\subsection{Solution} 
Solving the resulting system of partial differential equations yields the Kerr-like metric:

\begin{align}
ds^2 = & -\left(1 - \frac{r_q r}{\Sigma}\right)c^2 dt^2 \notag \\
& - \frac{2r_q r a \sin^2\theta}{\Sigma}c \, dt \, d\phi \notag \\
& + \frac{\Sigma}{\Delta}dr^2 + \Sigma d\theta^2 \notag \\
& + \left(r^2 + a^2 + \frac{r_q r a^2 \sin^2\theta}{\Sigma}\right)\sin^2\theta \, d\phi^2
\end{align}

where:
\begin{equation}
    r_q = \frac{4\pi}{\epsilon_0} \frac{Qq}{m_0 c^2}, \quad
    \Sigma = r^2 + a^2 \cos^2\theta, \quad
    \Delta = r^2 - r_q r + a^2, \quad
    a = \frac{J}{Qc}
\end{equation}

Here, \( Q \) is the total charge of the source, \( J \) is its angular momentum, and \( q \) and \( m_0 \) are the charge and rest mass of the test particle, respectively.

\subsection{Final Metric} 
This Kerr-like metric describes the geometry experienced by a test particle with charge \( q \) and mass \( m_0 \) near a rotating, spherically symmetric charge distribution with total charge \( Q \) and angular momentum \( J \). Key differences from the Kerr metric in general relativity are as follows:

\begin{itemize}
    \item \( r_q \) depends on the charge-to-mass ratio of the test particle
    \item \( a \) is interpreted as \( \frac{J}{Qc} \) rather than \( \frac{J}{Mc} \)
\end{itemize}

These differences reflect the electromagnetic nature of interactions in the pseudo-curvature framework.


\begin{thebibliography}{}
\expandafter\ifx\csname natexlab\endcsname\relax\def\natexlab#1{#1}\fi
\providecommand{\url}[1]{\href{#1}{#1}}
\providecommand{\dodoi}[1]{doi:~\href{http://doi.org/#1}{\nolinkurl{#1}}}
\providecommand{\doeprint}[1]{\href{http://ascl.net/#1}{\nolinkurl{http://ascl.net/#1}}}
\providecommand{\doarXiv}[1]{\href{https://arxiv.org/abs/#1}{\nolinkurl{https://arxiv.org/abs/#1}}}

\bibitem[{Burns {et~al.}(2024)Burns, Daniel, Alexander, \& Dressel}]{Burns_2024}
Burns, L., Daniel, T., Alexander, S., \& Dressel, J. 2024, Quantum Studies: Mathematics and Foundations, 11, 27–67, \dodoi{10.1007/s40509-024-00317-8}

\bibitem[{{Drill}(1939)}]{Drill}
{Drill}, H.~T. 1939, Physical Review, 56, 184, \dodoi{10.1103/PhysRev.56.184}

\bibitem[{Einstein(1915)}]{Einstein}
Einstein, A. 1915, Sitzungsber. Preuss. Akad. Wiss. Berlin, 778

\bibitem[{Feynman(1949)}]{Feynman}
Feynman, R.~P. 1949, Phys. Rev., 76, 769, \dodoi{10.1103/PhysRev.76.769}

\bibitem[{Kaluza(1921)}]{Kaluza1921}
Kaluza, T. 1921, Sitzungsber. Preuss. Akad. Wiss. Berlin (Math. Phys.), 1921, 966

\bibitem[{{Kennedy} \& {Thorndike}(1931)}]{Kennedy}
{Kennedy}, R.~J., \& {Thorndike}, E.~M. 1931, Proceedings of the National Academy of Science, 17, 620, \dodoi{10.1073/pnas.17.11.620}

\bibitem[{Klein(1926)}]{Klein1926}
Klein, O. 1926, Zeitschrift f{\"u}r Physik, 37, 895

\bibitem[{Schwinger(1948)}]{Schwinger}
Schwinger, J. 1948, Phys. Rev., 73, 416, \dodoi{10.1103/PhysRev.73.416}

\bibitem[{Wanas \& Ammar(2010)}]{Wanas2010}
Wanas, M.~I., \& Ammar, S.~A. 2010, Modern Physics Letters A, 25, 2883, \dodoi{10.1142/S0217732310032883}

\bibitem[{Weinberg(1995)}]{Weinberg_1995}
Weinberg, S. 1995, The Quantum Theory of Fields (Cambridge University Press)

\bibitem[{Weyl(1918)}]{Weyl1918}
Weyl, H. 1918, Sitzungsber. Preuss. Akad. Wiss., 465

\bibitem[{{Kerr}(1963)}]{kerr}
{Kerr}, R.~P. 1963, \prl, 11, 237, \dodoi{10.1103/PhysRevLett.11.237}

\bibitem[{{Schwarzschild}(1999)}]{schwarzschild1999gravitationalfieldmasspoint}
{Schwarzschild}, K. 1999, arXiv e-prints, \href{https://arxiv.org/abs/physics/9905030}{{physics/9905030}}


\end{thebibliography}
\end{document}